\documentclass[
    twocolumn,
	prd,
	amssymb,
	preprintnumbers,superscriptaddress,
	nofootinbib]{revtex4-1}

\pdfoutput=1

\usepackage{graphicx}
\usepackage{enumitem}
\usepackage{latexsym}
\usepackage{amsfonts}
\usepackage{amssymb}
\usepackage{xcolor}
\usepackage[export]{adjustbox}
\usepackage{amsmath}
\usepackage[thinlines]{easytable}
\usepackage{slashed}
\usepackage{dcolumn}
\usepackage{verbatim}
\usepackage{float}
\usepackage{multirow}
\usepackage{xspace}
\usepackage[normalem]{ulem}
\usepackage[
pdfauthor={Jeremy Sakstein}]{hyperref}
\usepackage{tabularx}

\newcommand{\beq}{\begin{equation}}
\newcommand{\eeq}{\end{equation}}
\newcommand{\bea}{\begin{eqnarray}}
\newcommand{\eea}{\end{eqnarray}}

  \newcommand{\bL}{\left[} \newcommand{\bR}{\right]}

\newcommand{\refjnl}[1]{{\rm#1}}

\def\apj{\refjnl{ApJ}}                 
\def\apjs{\refjnl{ApJS}}               
\def\aap{\refjnl{A\&A}}                

\newcommand{\pd}[2]{\frac{\partial #1}{\partial #2}}
\newcommand{\pdd}[2]{\frac{\partial^2 #1}{\partial #2^2}}

\newcommand{\msun}{{\rm M}_\odot}

\newcommand{\pdt}[3]{\frac{\partial^2 #1}{\partial #2 \partial #3}}

\newcommand{\pa}{\partial}

\newcommand{\nbh}{N_{\rm BH}}
\newcommand{\mbh}{M_{\rm BH}}
\newcommand{\mbhmg}{M_{\rm BHMG}}

\DeclareRobustCommand{\okina}{%
  \raisebox{\dimexpr\fontcharht\font`A-\height}{%
    \scalebox{0.8}{`}%
  }%
}

\allowdisplaybreaks

\begin{document}

\title{Axion Instability Supernovae}
\author{Jeremy Sakstein} \email{sakstein@hawaii.edu}
\affiliation{Department of Physics \& Astronomy, University of Hawai\okina i, Watanabe Hall, 2505 Correa Road, Honolulu, HI, 96822, USA}
\author{Djuna Croon} \email{djuna.l.croon@durham.ac.uk}
\affiliation{Institute for Particle Physics Phenomenology, Department of Physics, Durham University, Durham DH1 3LE, U.K.}
\affiliation{TRIUMF, 4004 Wesbrook Mall, Vancouver, BC V6T 2A3, Canada}
\author{Samuel D.~McDermott}
\email{sammcd00@fnal.gov}
\affiliation{Fermi National Accelerator Laboratory, Batavia, IL USA}

\date{\today}

\begin{abstract}
New particles coupled to the Standard Model can equilibrate in stellar cores if they are sufficiently heavy and strongly coupled.
In this work, we investigate the astrophysical consequences of 
such a scenario for massive 
stars by incorporating new contributions to the equation of state
into a state of the art stellar structure code. We focus on axions in the ``cosmological triangle'', a region of parameter space with $300{\rm\,keV} \lesssim m_a \lesssim 2$ MeV, $g_{a\gamma\gamma}\sim 10^{-5}$ GeV$^{-1}$ that is not presently excluded by other considerations. We find that for axion masses $m_a \sim m_e $, axion production in the core drives a new stellar instability that results in explosive nuclear burning that either drives a series of mass-shedding pulsations or completely disrupts the star resulting in a new type of optical transient --- an \textit{Axion Instability Supernova}. We predict that the upper black hole mass gap would be located at $37\msun \le M\le 107\msun$ in these theories, a large shift down from the standard prediction, which is disfavored by the detection of the mass gap in the LIGO/Virgo/KAGRA GWTC-2 gravitational wave catalog beginning at $46_{-6}^{+17}\msun$. Furthermore, axion-instability supernovae are more common than pair-instability supernovae, making them excellent candidate targets for JWST. The methods presented in this work can be used to investigate the astrophysical consequences of any theory of new physics that contains heavy bosonic particles of arbitrary spin. We provide the  tools to facilitate such studies.
\end{abstract}

\preprint{IPPP/22/10}
\preprint{FERMILAB-PUB-22-118-T}

\maketitle

\section{Introduction}

The observation of gravitational waves from merging binary black holes by the LIGO/Virgo/KAGRA (LVK) collaboration has enabled the study of the population statistics of astrophysical black holes \cite{LIGOScientific:2021psn}. The black hole mass function (BHMF) is predominantly shaped by the processes governing the structure, evolution, and fate of massive stars, making it a sensitive probe of stellar structure theory \cite{Farmer:2019jed,Farmer:2020xne}. It is also a novel and powerful probe of new physics beyond the Standard Model.~The evolution of massive stars can be altered by modifications of gravity that arise in leading dark energy theories \cite{Straight:2020zke}, additional energy losses from new light particles \cite{Croon:2020ehi,Croon:2020oga,Sakstein:2020axg}, and energy injections from dark matter self-annihilation \cite{Ziegler:2020klg, Ellis:2021ztw}, all of which leave imprints on the BHMF that can be distinguished with gravitational wave data from LVK \cite{Baxter:2021swn}. New physics process can affect both the shape of the BHMF and the locations of features such as the upper black hole mass gap (BHMG), which allows competing theories to be distinguished. 

The BHMG is a feature in the BHMF that refers to the absence of astrophysical black holes with masses in the range $50\msun\lesssim M\lesssim 120\msun$. The physical origin of this feature is the so-called \textit{pair-instability}, a stellar instability due to electron-positron pair-production in the cores of massive stars, which have densities and temperatures that are sufficient to produce copious amounts of non-relativistic $e^-e^+$ pairs. Small contractions of the star result in temperature and density increases; a corresponding increase of pressure would be expected to counteract the contraction, but leads instead to an increase in the rate of $e^-e^+$ production \cite{1967ApJ...148..803R}. This causes a runaway contraction that is only halted by explosive oxygen ignition after the core temperature and density have become sufficiently high. The resulting explosion unbinds the star --- a process referred to as a \textit{pair-instability supernova} (PISN) --- leaving no black hole (BH) remnant. A recent analysis \cite{Baxter:2021swn} of the LVK GWTC-2 catalog has determined the location of the lower edge of the BHMG to be $M_{\rm BHMG}=46_{-6}^{+17}\msun$ ($M_{\rm BHMG}=54_{-6}^{+6}\msun$ if GW190521, a potential outlier, is excluded from the analysis).

In a recent publication \cite{Croon:2020oga}, we identified the existence of a new stellar instability triggered by heavy new particles. If the new particle couples to the Standard Model (SM) with sufficient strength then it can be produced in the cores of massive stars through SM interactions.~{Heavy particles are especially interesting because they cannot be produced in well-studied stars with lower masses.}~If the particles are subsequently unable to free-stream out of the star, they can come into equilibrium with the stellar plasma.~Rather than acting as a new source of energy loss, as is the case with light particles, heavy particles alter the equation of state (EOS) of the thermal plasma \cite{1967ApJ...148..803R}. This process is analogous to the $e^-e^+$ pair-instability. The heavy new particles are produced with non-relativistic velocities, and therefore rob the star of pressure support and act to destabilize the star, resulting in a similar runaway contraction. In \cite{Croon:2020oga}, we demonstrated the existence of such an instability by calculating the equation of state of a gas of coupled ions, photons, electrons, positrons, and heavy new particles.~This is sufficient to determine the existence of the instability, but calculating its effect on the structure, evolution, and fate of massive objects requires detailed stellar evolution simulations.

In this paper, we present the  results of incorporating heavy new particles into the EOS of stellar matter on the fates of massive stars. We have modified the stellar structure code MESA \cite{Paxton:2010ji,Paxton:2013pj,Paxton:2015jva,Paxton:2017eie,2019ApJS..243...10P} to include an additional sector of heavy particles in the equation of state. We focus on axions with masses $m_a\sim 500$ keV coupled to photons with coupling strength $g_{a\gamma\gamma}\sim 10^{-5}$ GeV$^{-1}$. This corresponds to the ``cosmological triangle" --- a small region in the $\{ g_{a \gamma \gamma}, m_a \}$ parameter space that is not currently excluded by other probes of axions \cite{Carenza:2020zil,Dolan:2021rya}. We find that stars with initial mass $M\gtrsim  48\msun$ undergo explosive oxygen burning in the star's core. The explosion is so violent that it unbinds the entire star, leaving no BH remnant.~Thus, we predict the existence of a new optical transient:~an \textit{Axion Instability Supernova} (AISN). In stars with initial masses in the range $38\msun\lesssim M\lesssim48\msun$ the instability drives a series of mass-shedding pulsations similar to the pulsational pair instability supernova (PPISN) process driven by the pair-instability. We therefore refer to this process and the associated novel optical transient as a \textit{Pulsational Axion Instability Supernova} (PAISN). Stars with mass $M\lesssim38\msun$ avoid the instability and core-collapse to form BHs with masses nearly identical to the initial mass. The upper black hole mass gap would therefore be located at $M_{\rm BHMG}\approx 37\msun$, in stark constant to the SM prediction $M_{\rm BHMG}\approx 50\msun$ \cite{Farmer:2020xne,Baxter:2021swn}. We also find that the the upper edge of the BHMG is located at $107\msun$, significantly lower than the SM prediction of $133\msun$, making it more likely that LVK can detect BHs on the far side of the gap.
The results above are robust to variations in metallicity and the ${}^{12}{\rm C}(\alpha,\gamma)^{16}{\rm O}$ rate. The detection of the upper BHMG feature in the LVK GWTC-2 gravitational wave catalog beginning at $M_{\rm BHMG}=46_{-6}^{+17}\msun$ \cite{Baxter:2021swn} therefore disfavors this axion mass. 

This paper is organized as follows.~In section \ref{sec:triangle} we discuss the cosmological triangle for axions coupled to photons and demonstrate that particles in this region can equilibriate in stellar cores.~In section \ref{sec:stellar_modeling} we describe the stellar structure code used for our simulations, derive the equation of state for heavy new particles, and explain its implementation into MESA. In section \ref{sec:results} we present our results. We conclude in section \ref{sec:concs}.

\section{Equilibration in the Cosmological Triangle}
\label{sec:triangle}
The methods presented in this work can be applied to any bosonic particle (of arbitrary multiplicity) which equilibrates in post-helium burning stellar cores such as axions, chameleons, symmetrons,  dilatons, or hidden photons. An important application, to which we devote our analysis in the rest of this paper, is axion-like particles (henceforward axions) in the so-called cosmological triangle (see, e.g., \cite{Carenza:2020zil,Dolan:2021rya}):~an open region in the $\{ g_{a \gamma \gamma}, m_a \}$ parameter space bordered by constraints from beam-dump experiments, supernova cooling constraints, and stellar bounds.~{The cosmological triangle received its name from its shape, and the fact that cosmology-dependent constraints can be deduced {given additional assumptions on exotic contributions to} $\Delta N_{\rm eff} $} \cite{Depta:2020wmr}.
This region of parameter space has recently been criticized based on limiting the energy deposition in the mantle and the outer envelopes of low-energy supernovae \cite{Caputo:2021rux, Caputo:2022mah}. These limits are suggestive that new optical phenomena are possible in this range of parameter space; however, until observational signatures of this type of energy deposition are better characterized, it is important to pursue additional probes of this parameter space.~In this subsection we show that axions in the cosmological triangle will indeed equilibrate on timescales relevant to massive star evolution, and in subsequent sections we work out additional observational clues to the presence of this new kind of particle.

There are two process relevant for heavy axion production in the mass and coupling range considered in this work --- photon conversion (the Primakoff process) and photon coalescence \cite{DiLella:2000dn,Carenza:2020zil,Lucente:2022wai}. The {production} rate for axions {by the process of photon conversion is} \cite{DiLella:2000dn,Dolan:2021rya}
\begin{equation} \label{eq:axion-prod-rate}
    \begin{split}
    &\Gamma_{ \gamma \to a} =
    \frac{g_{a\gamma\gamma}^2Tk_s^2}{32\pi}\frac{k}{\omega}
    \Bigg\{\frac{[(k+p)^2+k_s^2][(k-p)^2+k_s^2]}{4kpk_s^2} \times \\
    &\ln\bigg[\frac{(k+p)^2+k_s^2}{(k-p)^2+k_s^2}\bigg]
    -\frac{(k^2-p^2)^2}{4kpk_s^2}\ln\bigg[\frac{(k+p)^2}{(k-p)^2}\bigg]-1
    \Bigg\},
\end{split}
\end{equation}
where $k=|\vec{k}| $ is the photon momentum, $\omega$ its energy, $p=|\vec{p}|$ is the axion momentum, $T$ is the plasma temperature, and $k_s$ describes plasma screening effects:
\begin{equation}
    k_s^2=\frac{4\pi\alpha}{T}\frac{\rho}{m_u}\bigg(Y_e + \sum_jZ_j^2Y_j\bigg),
\end{equation}
where $ Y_e$ is the number of electrons per baryon, $Y_j$ is the number per baryon of nuclear species with charge $Z_j$, and $m_u $ is the atomic mass unit. The {production} rate for axions {by the process of photon coalescence is (see e.g. \cite{Cadamuro:2010cz,Cadamuro:2011fd,Bastero-Gil:2021oky})}
\begin{equation}
    \Gamma_{a\gamma\gamma}=\frac{ma^2-4\omega_p^2}{m_a^2}\left(\frac{m_a}{E}\right)\left[1+\frac{2T}{p}\ln\left(\frac{1-e^{-(E+p)/2T}}{1+e^{-(E-p)/2T}}\right)\right],
\end{equation}
where the axion decay rate is given by
\begin{equation}
    \Gamma_{a\gamma\gamma}=\frac{g_{a\gamma\gamma}, m_a^3}{64\pi}\quad\textrm{and}\quad \omega_p^2=\frac{4\pi\alpha n_e}{T},
\end{equation}
where the latter is the plasma frequency ($n_e$ is the electron number density).

The equilibration time can be estimated as in \cite{Croon:2020oga} as $\tau_{\rm eq} \simeq \left( \sum \Gamma_{a\rightarrow\gamma} \right)^{-1}$, where the sum runs over all processes that allow the absorption of axions, since the rate of absorptive processes must equal the rate of production when in equilibrium. The rate of core temperature evolution in our simulation can be estimated as $\Gamma_{\rm evol} = \dot{T}/T$ where the dot indicates a derivative with respect to time.~We can safely assume an equilibrium population of axions if $ \sum\Gamma_{ \gamma \to a} > \Gamma_{\rm evol}$ before the onset of the PISN or PPISN.~Given that the decay rate alone, estimated as $\Gamma_{\rm decay} \sim g_a^2 m_a^3$, is of order microseconds for the mass and coupling considered in this work, this condition is satisfied in the cores of the stars that we simulate, except possibly during the pulsations. We therefore assume an equilibrium distribution during quasi-static stellar burning and make the conservative choice to disable our modifications to the EOS at the onset of the pulsational stage of evolution.

\section{Stellar Modelling}
\label{sec:stellar_modeling}

Our simulations were performed using the publicly available stellar structure code MESA version 12778 \cite{Paxton:2010ji,Paxton:2013pj,Paxton:2015jva,Paxton:2017eie,2019ApJS..243...10P}. Our modifications to MESA as well as a Mathematica code to produce equations of state for user-supplied masses and spins are publicly available at the following URL \href{https://zenodo.org/record/6347632}{https://zenodo.org/record/6347632} \cite{jeremy_sakstein_2022_6347632}.

\subsection{Input Physics}

MESA is a one-dimensional (meaning that it assumes spherical symmetry) hydrostatic code. It is equipped with a Harten–Lax–vanLeer–Contact (HLLC) hydrodynamical solver that can be switched on to simulate departures from hydrostatic equilibrium such as pulsations and shocks \cite{2019ApJS..243...10P}. Our simulation of the pulsations and stellar explosions due to new particles is identical to the implementation described in \cite{2019ApJS..243...10P,Marchant:2018kun,Farmer:2019jed,Croon:2020oga,Straight:2020zke}.  

Relevant prescriptions for our simulations are as follows. We treat convection according to the Cox prescription for mixing length theory \cite{1968pss..book.....C} with efficiency parameter $\alpha_{\rm MLT}=2.0$. Semi-convection is modeled according to  \cite{1985A&A...145..179L} with efficiency parameter $\alpha_{\rm SC}=1.0$. Convective overshooting is described using an exponential profile parameterized by $f_0$, which sets the point inside the convective boundary where overshooting begins, and $f_{\rm ov}$, which determines the scale height of the overshoot. We set $f_0=0.005$ and $f_{\rm ov}=0.01$. Our prescription for mass loss due to stellar winds  follows that of \cite{Brott:2011ni} with the wind efficiency parameter (clumping parameter) $\eta$ fixed to $0.1$. Finally, we use the MESA default nuclear burning rates (these are a mixture of the {NACRE} \cite{Angulo:1999zz} and {REACLIB} \cite{2010ApJS..189..240C} tables) with the one exception of the $^{12}{\rm C}(\alpha,\gamma)^{16}{\rm O}$ reaction. Previous studies have found this to be the most important rate for determining the location of the upper BHMG \cite{Farmer:2019jed,Farmer:2020xne,Mehta:2021fgz} so we use the most recent results from \cite{deBoer:2017ldl}. In particular, we use their median values (changing this by $\pm1\sigma$ changes the location of the SM BHMG by $\pm2.5\msun$ \cite{Farmer:2020xne,Baxter:2021swn}). 

We derive the final black hole mass by simulating helium cores from the zero age helium branch (ZAHB) until either core collapse or PISN/AISN. The black hole mass is calculated as the mass of  the layers expanding with velocity less than the escape velocity. The inlists given in the reproduction package for this paper are identical to those used to generate our results.

\subsection{Equation of State for Heavy Bosonic Particles}

We assume that the new particle is bosonic with $g=2s+1$ (with $s$ the spin) degrees of freedom and further assume that it {has no gauge quantum numbers}, implying that it has zero chemical potential. This assumption can easily be relaxed in more complicated models.~The contribution to the EOS can then be calculated by integrating over the distribution. Here we assume a thermal equilibrium Bose-Einstein distribution. We leave a detailed study of the thermalization processes for future work. Our procedure for calculating the EOS for heavy new particles and its implementation into MESA is described in Appendix \ref{sec:App_mesa_EOS}.

We assume that all of the axions that are produced by the process described in Eq.~\ref{eq:axion-prod-rate} remain trapped in the stellar core. This is only an approximation:~some of the axions produced from photon conversion will be generated with velocities exceeding the escape velocity from the star. However, for the parameter values we choose, such particles should be able to scatter with SM particles on their way out of the star, and in doing so will on average lose energy to the gravitationally bound stellar material, which will lead them to eventually become trapped and populate the thermal phase space as assumed here. The process by which they scatter and become bound could lead to transport of internal stellar energy (see e.g. \cite{Gould:1989ez,Gould:1989hm,Raffelt:1988rx,Raen:2020qvn,Lucente:2022wai}). The energy transport and subsequent redistribution in the core reduces the temperature gradient and hence the amount of convection.~Convective mixing of $^{12}{\rm C}$ inhibits the explosive burning of Oxygen \cite{Farmer:2020xne} so additional energy transport by axions is likely to result in stronger pulsations and the onset of instability at lower ZAHB mass. Full numerical studies are needed to confirm this. For this reason, we believe that a full, out-of-equilibrium simulation of the dynamics of the axion population in the star will be an interesting topic to study in greater detail in future work.~Nevertheless, our study is of value for the limiting cases of very tightly coupled new particles or modifications of the EOS induced by modified gravity or new couplings between SM particles.

\section{Results}
\label{sec:results}

\begin{figure}[t]
    \centering
    \includegraphics[width=.49\textwidth]{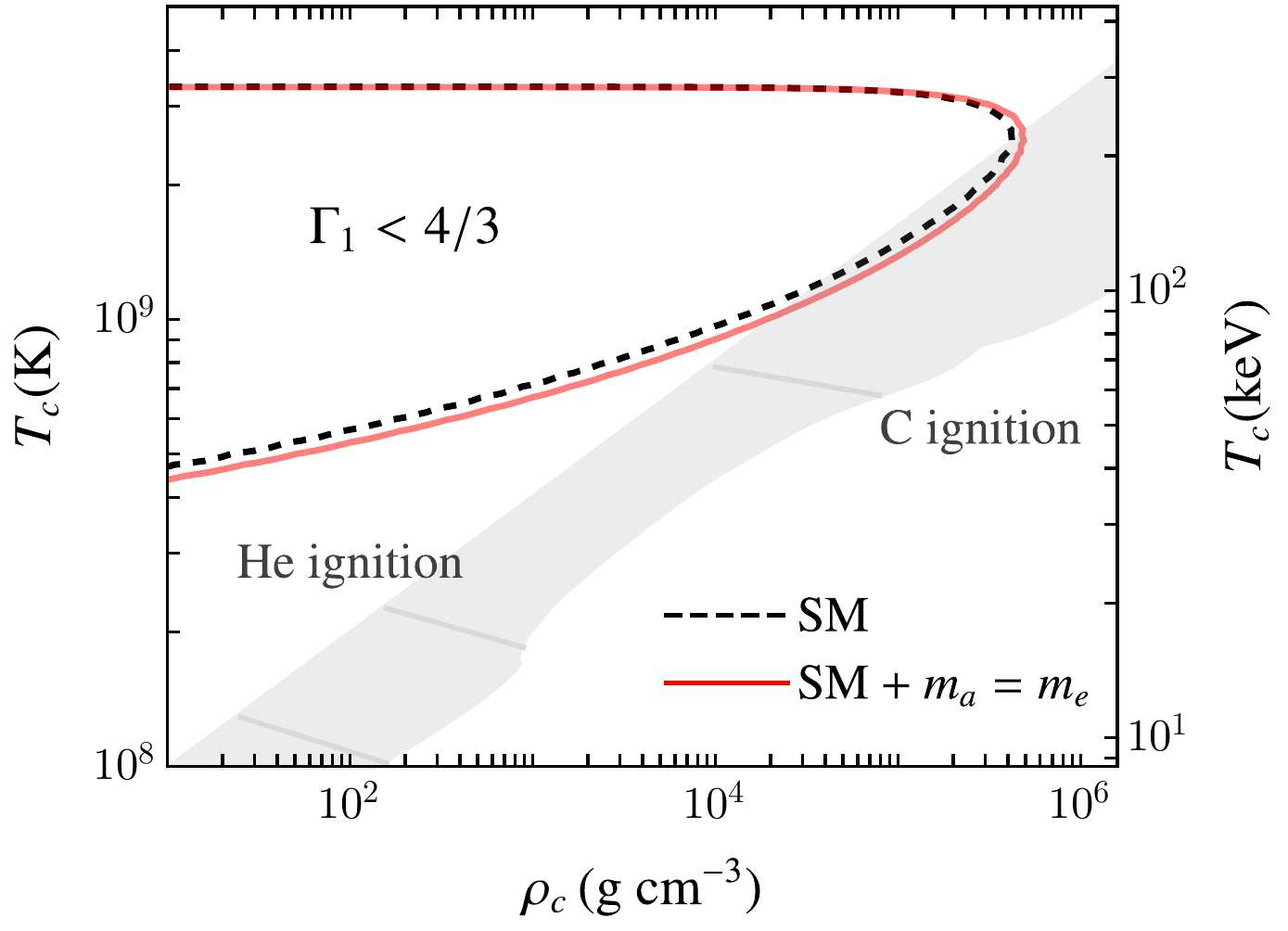}
    \caption{The central temperatures and densities where the pair-instability (black dashed line) and axion-instability (red solid line) are encountered. The axion mass was taken to be equal to the electron mass. The gray region indicates the approximate location of typical massive star evolutionary tracks.
    }
    \label{fig:instability}
\end{figure}

{We report here on the impact of an axion with mass $m_a=m_e=511$ keV --- near the lower edge of the cosmological triangle. The effects of less-massive axions will be even more dramatically visible than an electron-mass axion. Intriguingly, we have verified that if the axion has a mass $m_a \gtrsim 2m_e$ then stars will encounter the PISN {\it before} the AISN, thus recovering Standard Model-like behavior anticipated with no axion at all. Below, we comment on the significance of these relations for LVK observations.}

\subsection{Physics of the Axion Instability}
{The inclusion of axions with masses equal to electrons exacerbates pair-instability.}~The temperatures and densities where the axion-instability manifests are shown in Figure \ref{fig:instability}. 
This figure demonstrates that adding a bosonic particle with the same mass as electrons implies an extension of the region with $ \Gamma_1 = 4/3$, where the stellar core is unstable.~We therefore expect lower mass objects to experience mass-shedding pulsations (PAISN) and complete disruption (AISN) compared with the SM prediction.~We also expect that these events will be more violent than the PPISN and PISN predicted by the SM. For $m_a \lesssim 2m_e$, axion production is efficient under similar conditions as $e^-e^+$ production, such that the effects of the pair-instability and axion-instability compound, resulting in a smaller value of $\Gamma_1$ at fixed temperature and density and therefore a stronger inward contraction when the instability is first encountered.~This stronger contraction can only be countered by a more violent oxygen explosion.

A second consequence of the compounding of the axion and pair instabilities is that lighter BHs can be formed at the upper edge of the upper black hole mass gap. The BHs reappear at high masses ($\mbh=133\msun$ in the SM) because the core temperatures and densities are sufficient for photodisintegration reactions to occur. These result in their own independent instability because the temperature and density increase due to small contractions further photodisintegrates the heavy elements rather than raising the pressure to counteract the contraction. Photodisintegrations compound with the pair-instability to lower $\Gamma_1$ further below $4/3$ to the point where the oxygen ignition is not sufficient to reverse the contraction and core collapse ensues. One would therefore expect a similar phenomenon resulting from the compounding of the axion and pair instabilities. This compounding happens at smaller temperatures and densities than those required for photodisintegration reactions, implying that the oxygen explosion will not be sufficient to reverse the implosions of some lower mass objects.

\begin{figure}[t]
    \centering
    \includegraphics[width=0.49\textwidth]{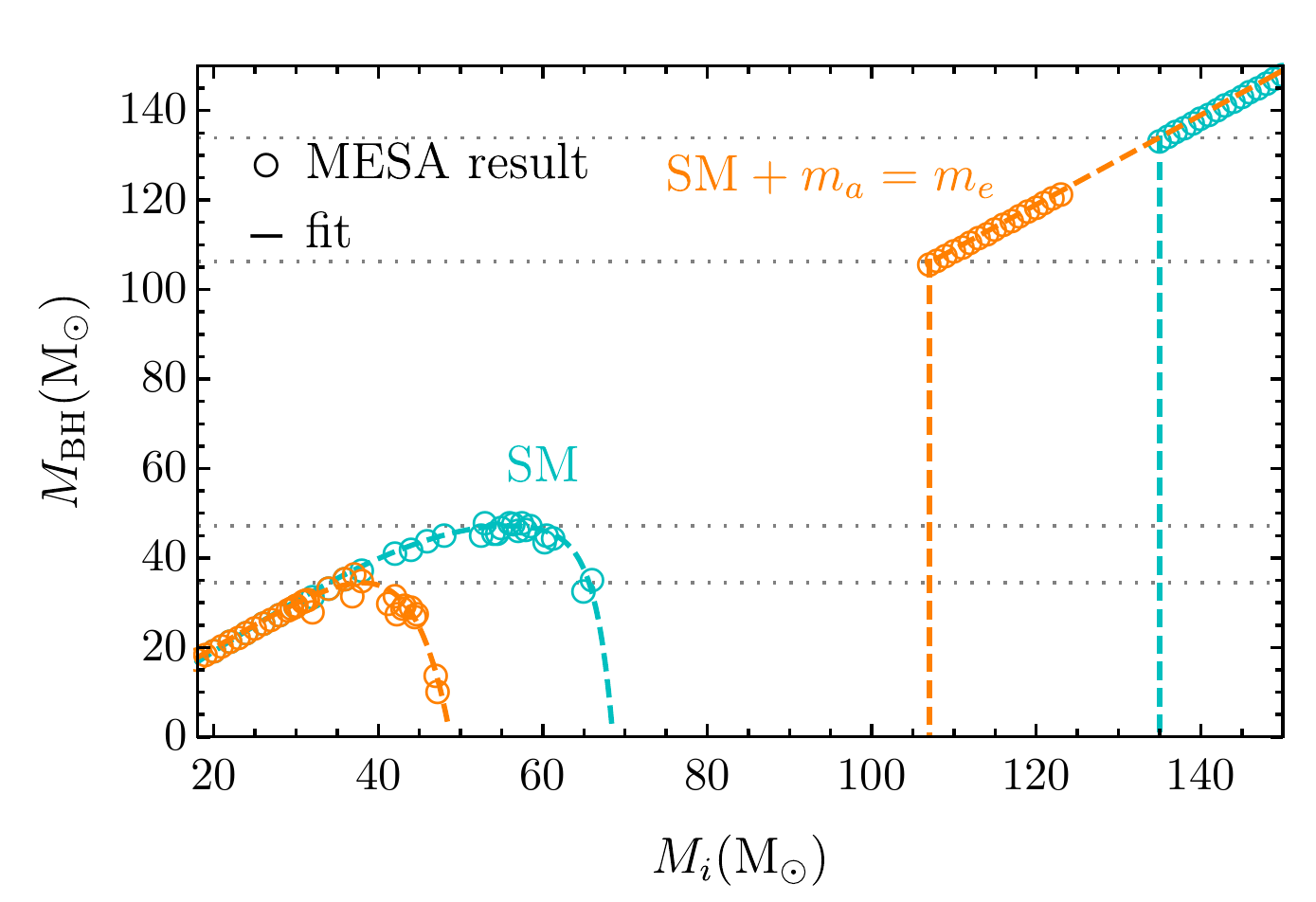}
    \caption{The black hole mass as a function of initial mass in both the SM and when axions with mass $m_a=m_e$ are included in the EOS. These results are for $Z=10^{-5}$. Each open circle corresponds to the result of an individual MESA simulation. }
    \label{fig:upper}
\end{figure}

\begin{figure*}[t]
    \centering
    \includegraphics[width=.99\textwidth]{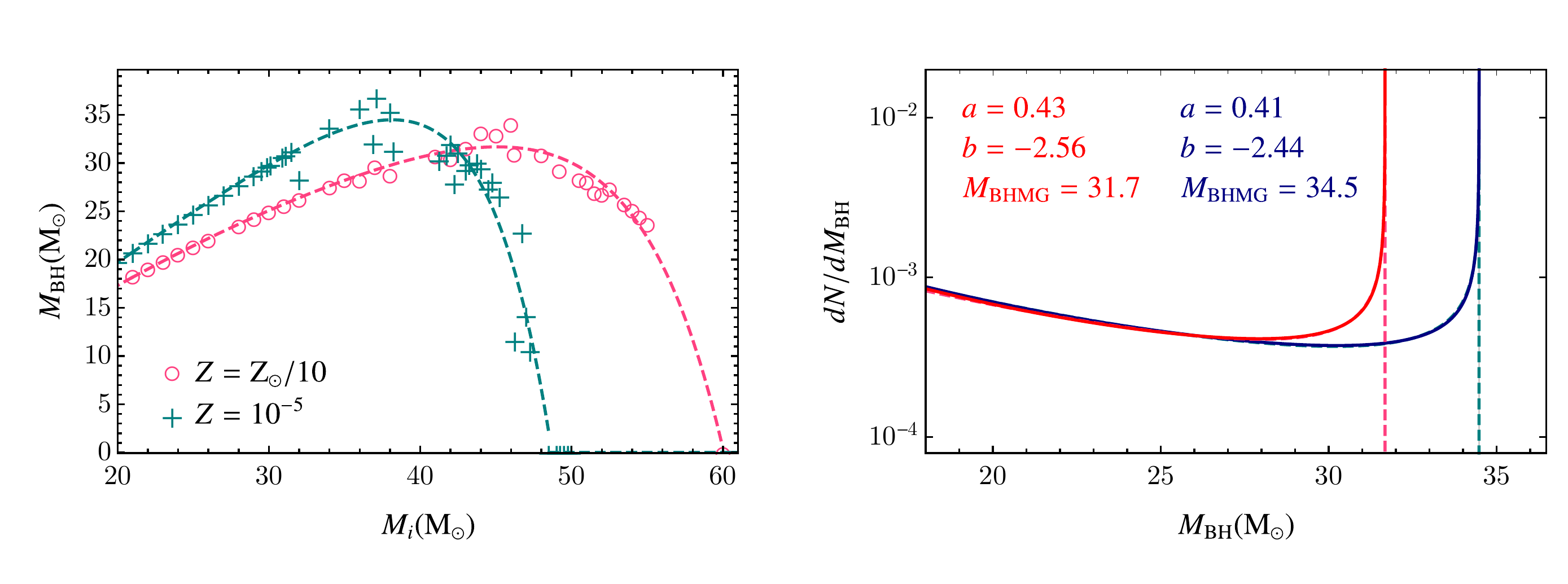}
    \caption{First generation black hole masses as a result of (P)AISN. \emph{Left:} simulation results and numerical fit. \emph{Right:} black hole mass function found using the fit in the left panel and the assumption that the initial mass function of the stellar population is a powerlaw with index $ -2.4$. It is seen that the value of $ M_{\rm BHMG}$ is lower than the value found in the Standard Model (see \cite{Baxter:2021swn}) by more than $10 \, \rm M_\odot$.}
    \label{fig:mfme}
\end{figure*}

\subsection{Location of the Upper Black Hole Mass Gap}

The results of our simulations are presented in figure \ref{fig:upper} where we plot the BH mass as a function of initial helium core mass. The metallicity was taken to be $Z=10^{-5}$ corresponding to population-III stars.~The qualitative effects described above are evident. The pulsations and total disruptions (AISN) indeed begin at lower masses compared with the SM, and stars that would have experienced pulsations are instead instead totally disrupted, leaving no BH remnant.~Furthermore, the BHs reappear at smaller masses compared with the SM. Our simulations predict that, when electron-mass axions are included in the EOS, the lower edge of the upper BHMG is located at $37\msun$ and the upper edge of the BHMG is located at $\sim 107\msun$ (compared to $\mbh = 133 \msun$ in the SM). Thus, we predict that the BHMG is narrowed from $86\msun$ to $71\msun$.

We have run additional grids with $Z=Z_\odot/10$ (corresponding to population-II stars) where the solar metallicity is $Z_\odot=0.0142$ and where the rate of the $^{12}{\rm C}(\alpha,\gamma)^{16}{\rm O}$ is varied by $\pm1\sigma$ from its median value. In all cases we find $\lesssim4\msun$ changes in the location of the lower edge of the BHMG. Thus, our predictions 
are similarly robust as the predictions of the PISN \cite{Farmer:2019jed}.

\subsection{Black Hole Population Signatures}

The results of our simulations are shown in the left panel of Fig.~\ref{fig:mfme}. To understand the function $\mbh(M_i)$, we perform a numerical fit to a continuous, parametric function which can reproduce the shape of the curve: a constant, two power-laws with arbitrary coefficients to model both the BHs unaffected by pair-instability and PAISN BHs, and an exponential fall-off capturing AISN, as introduced in \cite{Baxter:2021swn}. In the right panel, we perform a fit to the {black hole} mass function introduced in \cite{Baxter:2021swn}: 
\begin{equation}
    \frac{d \nbh^{\rm(1g)}}{d \mbh} \! \propto \! \mbh^b \! \bL 1 \! + \! \frac{2 a^2 \mbh^{1/2}
(\mbhmg - \mbh)^{a-1}}{ \mbhmg^{a-1/2}} \! \bR \! .
\label{eq:mf}
\end{equation}
It is seen that the proposed mass function approximates the shape of this MESA-fit mass function well with just three parameters. The best fit parameters are also given in Fig.~\ref{fig:mfme}. Strikingly, it is seen that the model \eqref{eq:mf} indicates a value of the BHMG of $ \mbhmg = 34.5\,  \msun$ for $Z=\rm 10^{-5}$ ($ \mbhmg = 31.7\,  \msun$ for $Z=\rm Z_\odot/10$). Fitting this function to the LIGO/Virgo/KAGRA GWTC-2 catalog yields a robust measurement of $M_{\rm BHMG}=46_{-6}^{+17}\msun$ ($54_{-6}^{+6}\msun$ if GW190521 is excluded from the analysis) \cite{Baxter:2021swn}, which seemingly disfavors the possibility of AISN. However, for axion masses $m_a \gtrsim 2$ MeV, we have determined that stars will encounter the PISN instead of the AISN. In this case, the functions $\mbh(M_i)$ and $d\nbh(\mbh)$ are the same as if there is no axion at all. Thus, we suggest the intriguing possibility that a spatially varying axion mass, such as could be generated by locally varying CP-violation in different host galaxies \cite{OHare:2020wah}, could lead to two distinct peaks in the black hole mass population. Finding a complete, cosmologically viable model including this physical effect is an important target for further research. 

Black holes beyond the upper edge of the upper BHMG are currently outside the detection range of LIGO/Virgo/KAGRA, but planned upgrades to their sensitivities and the addition of planned future interferometers to the network (e.g. LIGO-India) will enable the location of the upper edge to be measured with percent-level precision \cite{Ezquiaga:2020tns}. Future detectors such as Cosmic Explorer (CE), Einstein Telescope (ET), and LISA will also be able to measure the location of the upper edge. It is therefore possible that the location of the upper edge will be measured within the next decade, at which point it could be used to place additional bounds on heavy axions.

\subsection{Optical Signatures}

Before concluding, we briefly comment on the possibility of directly detecting AISN using optical observations. AISN will be more common than PISN since their progenitors have lower masses, and lower mass stars are more numerous due to the negative slope of the stellar initial mass function (IMF).

Additionally, we find that at fixed progenitor mass AISN are brighter than PISN owing to the more violent oxygen explosion needed to counteract the stronger contraction. We plot the luminosity of a $52\msun$ and $70\msun$ AISN alongside a $70\msun$ PISN in figure \ref{fig:optical}. Evidently, a $52\msun$ AISN is brighter than a $70\msun$ PISN, whereas the $ 70 \msun$ AISN is brighter by more
than an order of magnitude. JWST will therefore see more AISN than PISN, making AISN excellent  targets. It would be interesting to use population synthesis and semi-analytic N-body codes to predict the AISN rate, and to use radiation hydrodynamics codes to predict their light curves. Such undertakings are beyond the scope of this work, but would provide additional detection channels.  

\section{Summary and Outlook}
\label{sec:concs}

In this work we have explored the effects of heavy axions in an unconstrained region of parameter space --- the cosmological triangle --- on the structure, evolution, and fate of massive stars. At these masses ($m_a\sim m_e$), these particles  can remain in the star and equilibrate with the stellar plasma, leading to a new instability akin to the pair-instability. We have explored the ultimate effects of this instability by incorporating heavy axions into the equation of state module of the stellar structure code MESA.

Our simulations predict that the axion-instability results in a contraction that ultimately unbinds the entire star. Thus we predict the existence of a new optical transient, an axion-instability supernova.

\begin{figure}[t]
    \centering
    \includegraphics[width=0.49\textwidth]{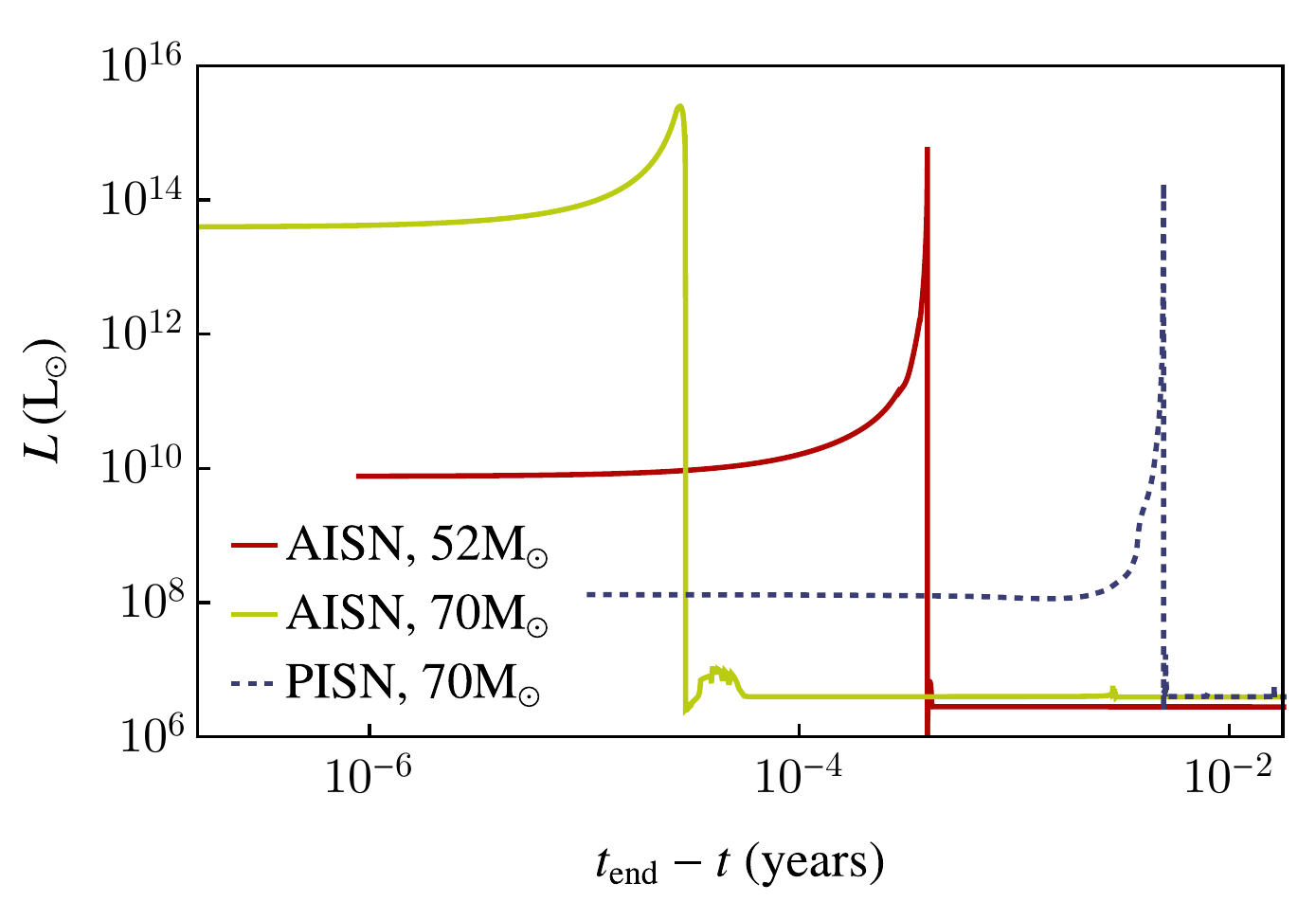}
    \caption{Luminosity of AISN and PISN.
    }
    \label{fig:optical}
\end{figure}
We investigated the axion-instability supernovae induced by an axion of mass $m_a=m_e$. We explored the impacts on the upper black hole mass gap and found that both the upper and lower edges are shifted to smaller masses. In particular, the lower edge is located at $37\msun$, compared to the SM prediction of $49\msun$, and the upper edge is located at $107\msun$, compared to the SM prediction of $133\msun$, as shown in Fig.~\ref{fig:upper}. Thus, the width of the gap narrows. The location of the lower edge is shifted to lower masses because the axion-instability manifests at higher densities than the pair-instability and can therefore be encountered by lower mass objects. The location of the upper edge is shifted to lower masses because the axion and pair instabilities compound as a result of axions in the cosmological triangle having similar masses to the electron. The resulting contraction is stronger than in the SM and the oxygen explosion is not sufficient to reverse it and prevent core collapse. We have checked that these results are robust to variations in metallicity and the ${}^{12}{\rm C}(\alpha,\gamma)^{16}{\rm O}$ rate. We also determined that axion-instability supernovae disappears for axions of mass $m_a \gtrsim 2 $ MeV, reverting to the standard astrophysical prediction. 

We studied the signatures of AISN on black hole populations. In the case of the lower edge, we demonstrated that an analysis of the population statistics of BHs in the LVK gravitational wave catalogs can distinguish between the two models. This could provide a robust detection of heavy axions in the cosmological triangle by fitting the BH mass catalog to the fitting function in Eq.~\ref{eq:mf}, which enables extraction of the location of the lower edge of the mass gap and the sharpness of the peak of the BH mass spectrum. Although these results appear to be in tension with current data, if there is spatial variation in the axion mass, 
then this could lead to a ``smearing'' of the peak in the black hole mass function, or even a mass function with multiple peaks. Evidence in favor of this possibility will be interesting to search for in black hole population catalogs. Planned upgrades to the sensitivities of LVK and the inclusion of additional detectors within the next decade will enable a robust measurement of the location of the upper edge, at which point independent corroborations can be placed.

A central assumption we have made in our analysis is that the new particles fill out the entirety of the local thermal phase space. This is inaccurate if they are produced with velocities greater than the escape velocity and have a mean free path much larger than the core size. In this case, they would contribute to anomalous energy transport in the star. In light of the findings here, a suite of out-of-equilibrium, momentum-dependent simulations of such new particles will be valuable to conduct.~We have also restricted to a single, spin-0, CP-odd degree of freedom, motivated by the existence of the ``cosmological triangle'' for axions, but our methods can be used to investigate the effects of new bosons with arbitrary masses and  spins.

The MESA code used to generate our results is included in the reproduction package for this paper  \cite{jeremy_sakstein_2022_6347632}. Also included are the codes necessary to generate modifications to the MESA EOS due to bosons with user-supplied masses and spins. We encourage the use of these for further science studies of new heavy particles coupled to the SM.

Finally, we note that it would be interesting to predict the optical and neutrino signatures of axion-instability supernovae. Optical signatures could be predicted by generating synthetic light curves using photometric evolution codes, and neutrino signatures could be predicted using full three-dimensional simulations of axion-instability superovae. We found that AISN are brighter and occur more frequently than PISN {(we gave an example in Fig.~\ref{fig:optical})}, raising the tantalizing possibility that they may be directly detected by JWST. 

\section*{Software}
MESA version~12778, MESASDK version 20200325, Mathematica version 12.0.

\section*{Acknowledgements}

We are grateful to Andrea Caputo, Robert Farmer, Adam Jermyn, Pablo Marchant, Mathieu Renzo, Frank Timmes, and Edoardo Vitagliano for insightful discussions and answering our many MESA-related questions. Our simulations were run on the University of Hawai\okina i's high-performance supercomputer MANA. The technical support and advanced computing resources from University of Hawai\okina i Information Technology Services – Cyberinfrastructure, funded in part by the National Science Foundation MRI award \#1920304, are gratefully acknowledged. Fermilab is operated by Fermi Research Alliance, LLC under Contract No.~De-AC02-07CH11359 with the United States Department of Energy, Office of High Energy Physics.

\appendix

\section{New Particle EOS and Implementation into MESA} \label{sec:App_mesa_EOS}

The MESA EOS module requires 16 thermodynamic variables to be specified. Three of them --- the mean molecular weight per gas particle $\mu$, the mean number of free electrons per nucleon $\mu_e^{-1}$, and the ratio of the electron chemical potential to $k_B T$ (degeneracy parameter $\eta$) --- are not modified by the presence of new particles. The remaining 13 are listed in table \ref{tab:mesa_EOS}. In addition to these, derivatives of each of these quantities with respect to $\ln T$ and $\ln\rho$ must also be provided. The EOS (and derivatives) are critical for determining the MESA time-step so must be accurate to high precision.

The default MESA EOS at temperatures and densities relevant for post-main-sequence massive star evolution is the HELM EOS \cite{2000ApJS..126..501T}. Our procedure for including new particles in the EOS is to first call this EOS and then modify each of the quantities listed in table \ref{tab:mesa_EOS}. The MESA EOS uses density and temperature as input variables. Consequentially, the contribution of new particles to quantities defined as derivatives at constant $\rho$ and $T$ e.g. $c_V$ and $\chi_\rho$, as well as those with no derivatives e.g. pressure can be found by calculating the corresponding quantity for new particles and adding this to the MESA default value. Quantities defined as derivatives with other variables held constant e.g. $\Gamma_1$ (which involves constant entropy) and $c_P$ (which involves constant pressure), cannot be added in this manner. The reason for this is that the MESA default value is calculated at constant default quantity which, with the exception of $T$ and $\rho$, is altered by the presence of new particles. In these cases, one must calculate the relevant quantity using thermodynamic relations when all species are included simultaneously (visible matter and new particles). We have derived formulas that can be used to calculate these quantities from a combination the MESA default EOS quantities, which we denote using overbars, and the EOS variables that can be updated to include new particles by simply adding their contribution. These formulas are listed in the final column of table \ref{tab:mesa_EOS}. 

Our procedure is then to first calculate all of the EOS variables that can be found by adding the contribution from new particles, and then to use these to calculate the remaining quantities. This mandates that the EOS variables be calculated in a specific order since some non-additive EOS variables must be specified before others can be calculated. From top to bottom, the order in which the quantities appear in table \ref{tab:mesa_EOS} reflects the order in which they are calculated in our MESA code. Given any single row, quantities from previous rows may act as inputs to the formula but not quantities from subsequent rows. The starting point for this process is the gas pressure $P_{\rm g}$, specific internal energy $E$, and specific entropy $s$. All other quantities can be calculated from combinations of these three and further quantities derived therefrom. 

Assuming that the new particle is a spin-$s$ boson $\phi$ with mass $m_\phi$ and degeneracy $g_\phi=2s+1$ which is in thermal equilibrium with the stellar material, these three quantities are calculated analytically from first principles by integrating over the Bose-Einstein distribution as follows. We begin by defining 
\begin{align}
C&=\frac{1}{\pi^2}\left(\frac{m_\phi c}{\hbar}\right)^3,\quad\textrm{and}\\ \beta(T)&=\frac{m_\phi c^2}{k_B T},
\end{align}
where $m_\phi$ is the mass of the new particle. The relevant thermodynamic quantities are the pressure, density, internal energy, and specific entropy given respectively by
\begin{align}
P_\phi(\beta)&=m_\phi c^2C\left( \frac{g_\phi}{2}\right)H_1(\beta) \label{eq:pres_def} \\\rho_\phi(\beta)&=m_\phi C\left( \frac{g_\phi}{2}\right)H_2(\beta)\\
u_\phi(\beta)&=m_\phi c^2C\left( \frac{g_\phi}{2}\right)H_3(\beta) \\ s_\phi&=\frac{k_B C{\beta}}{\rho}\left( \frac{g_\phi}{2}\right)\left[H_1(\beta)+H_3(\beta)\right]\label{eq:s_def}
\end{align}
 where 
\begin{align}
H_1(\beta)&=\int_{\varepsilon=\beta}^\infty G\left(\frac{\varepsilon}{\beta}\right)B(\varepsilon)\frac{\mathrm{d}\varepsilon}{\beta}\\ H_2(\beta)&=\int^\infty_{\varepsilon=\beta} G' \! \left(\frac{\varepsilon}{\beta}\right) B(\varepsilon)\frac{\mathrm{d}\varepsilon}{\beta}\\
H_3(\beta)&=\int^\infty_{\varepsilon=\beta}\varepsilon \, G' \! \left(\frac{\varepsilon}{\beta}\right)B(\varepsilon)\frac{\mathrm{d}\varepsilon}{\beta^2}\\ B(\varepsilon) &=\frac{1}{e^{\varepsilon}-1}\\
G(x)&=\frac13(x^2-1)^{\frac32}.
\end{align}
The assumption that the phase space distribution of axions follows its thermal equilibrium form is nontrivial, since the production processes are athermal. Nevertheless, thermal broadening, redshifting, and rescattering will push the distribution towards thermalizing. Exploring the process of thermalization in detail is an interesting topic for future work.

The formulas for the EOS variables that we use in our MESA code are given in table \ref{tab:mesa_EOS}. The formulas for the derivatives of these variables with respect to temperature are given in table \ref{tab:derive_const_T} and the formula for the derivatives with respect to density are given in table \ref{tab:derive_const_d}. The derivatives are crucial for determining the MESA timesteps and cannot be omitted. In order to enable rapid computation and maintain high precision we have derived analytic fitting functions for $P_\phi$, $u_\phi$, and $s_\phi$, and their derivatives with respect to temperature and density. Specifically, we fit an eighth-order polynomial in $T$ to each formula (the $\rho$-dependence is simple and can be included analytically) and evaluate the formulas in the final columns of the tables in MESA using these fitting functions. A copy of the specific fitting formulas used in this work (corresponding to a $511$ keV scalar) is included with the reproduction package for this paper (see \href{https://zenodo.org/record/6347632}{https://zenodo.org/record/6347632} \cite{jeremy_sakstein_2022_6347632}). Our reproduction package includes a Mathematica script that generates these fitting formulas for arbitrary boson masses $m_\phi$ and degeneracies $g_\phi$ for immediate use.  

\setlength\extrarowheight{15pt}
\begin{table*}[h]
    \centering
    \begin{tabular}{|c|c |c |c| }\hline
      Quantity    & Definition & Units in MESA & Formula\\\hline  $P_g$ &  gas pressure   & ergs/${\rm cm}^3$ & $P_g = \bar{P}_g+P_\phi$ 
\\\vspace{-2.5ex} & & & \vspace{-2.5ex}\\\hline
      $\displaystyle E$ & specific internal energy  & ergs/g & $\displaystyle E=\bar{E}+\frac{u_\phi}{\rho}$\\\vspace{-2.5ex} & & & \vspace{-2.5ex}\\\hline
      $\displaystyle s$ & specific entropy  & ergs/g/K & $\displaystyle s=\bar s + s_\phi$\\\vspace{-2.5ex} & & & \vspace{-2.5ex}\\\hline
      $\displaystyle c_V$ &  $\displaystyle\left(\frac{\pa E}{\pa T}\right)_\rho$ & ergs/g/K & $\displaystyle \bar{c}_v-k_BC\frac{g_\phi}{2}\frac{\beta^2}{\rho}\frac{\mathrm{d}H_3(\beta)}{\mathrm{d}\beta}$\\\vspace{-2.5ex} & & & \vspace{-2.5ex}\\\hline
      $\displaystyle\chi_\rho$ & $\displaystyle\left.\frac{\partial \ln P}{\partial \ln \rho}\right\vert_T$ & none & $\displaystyle \frac{\bar P}{P}\bar\chi_\rho$\\\vspace{-2.5ex} & & & \vspace{-2.5ex}\\\hline
      $\displaystyle \chi_T$ & $\displaystyle\left.\frac{\partial \ln P}{\partial \ln T}\right\vert_\rho$ & none & $\displaystyle \frac{\bar P}{P}\bar\chi_T-k_BC\frac{g_\phi}{2}\beta^2\frac{T}{P}\frac{\mathrm{d}H_1(\beta)}{\mathrm{d}\beta}$\\\vspace{-2.5ex} & & & \vspace{-2.5ex}\\\hline
      \multirow{2}{*}{  $\displaystyle\left(\frac{\pa s}{\pa T}\right)_\rho$} & --- & ergs/g/K$^2$ & $\displaystyle\left(\frac{\pa \bar s}{\pa T}\right)_\rho-\frac{k_B C}{\rho}\frac{g_\phi}{2}\frac{\beta}{T}\left[H_1(\beta)+H_3(\beta)\right.$\\ & & & $\displaystyle\left.+\beta\left(\frac{\mathrm{d}H_1(\beta)}{\mathrm{d}\beta}+\frac{\mathrm{d}H_3(\beta)}{\mathrm{d}\beta}\right)\right]$\\\vspace{-2.5ex} & & & \vspace{-2.5ex}\\\hline
      $\displaystyle\left(\frac{\pa s}{\pa \rho}\right)_T$ & --- & ergs cm$^3$/g$^2$/K & $\displaystyle\left(\frac{\pa \bar s}{\pa \rho}\right)_T-\frac{s_\phi}{\rho} $\\\vspace{-2.5ex} & & & \vspace{-2.5ex}\\\hline
      $\displaystyle\left(\frac{\pa E}{\pa\rho}\right)_T$ & --- & ergs cm$^3$/g$^2$ & $\displaystyle\left(\frac{\pa \bar E}{\pa\rho}\right)_T-\frac{u_\phi}{\rho^2}$\\\vspace{-2.5ex} & & & \vspace{-2.5ex}\\\hline
      $\displaystyle\Gamma_3$ & $\displaystyle1+\left(\frac{\pa \ln T}{\pa \ln \rho}\right)_s$ & none & $\displaystyle1+\frac{X}{Y}$\\\vspace{-2.5ex} & & & \vspace{-2.5ex}\\\hline
      $\displaystyle\Gamma_1$ & $\displaystyle\left(\frac{\pa \ln P}{\pa \ln \rho}\right)_s$ & none & $\displaystyle\chi_\rho + \chi_T\frac{X}{Y}$\\\vspace{-2.5ex} & & & \vspace{-2.5ex}\\\hline
      $\displaystyle\nabla_{\rm ad}$ & $\displaystyle\left(\frac{\pa\ln T}{\pa \ln P}\right)_s$ & none & $\displaystyle\frac{X}{Y\Gamma_1}$\\\vspace{-2.5ex} & & & \vspace{-2.5ex}\\\hline
      $\displaystyle c_P$ & $\displaystyle\left( \frac{\partial h}{\partial T} \right)_P$ & ergs/g/K &  $\displaystyle\frac{c_V}{\chi_\rho}\Gamma_1$\\\hline
    \end{tabular}
    \caption{The 13 MESA EOS variables altered by new particles. We have defined the quantities $X=P\chi_T$ and $Y=\rho T c_v$. The specific enthalpy is $h=E+P/\rho$. The gas pressure is defined as the total of all sources of pressure except the radiation pressure $P_{\rm rad}=aT^4/3$ with $a$ the radiation constant. The total pressure is then $P=P_g+aT^4/3$. Quantities with an overbar are those returned by the MESA default EOS and do not include DM. The formulas in the final column are those used in our modified MESA code. }
    \label{tab:mesa_EOS}
\end{table*}

\setlength\extrarowheight{15pt}
\begin{table*}[h]
    \centering
    \begin{tabular}{|c|c|}\hline
       Quantity  & Derivative  \\\hline
       $\displaystyle \ln(P_{\rm g})$  & $\displaystyle\frac{\bar P_{\rm g}}{P_{\rm g}}\pd{\ln P_{\rm g}}{\ln T} + \frac{T}{P_{\rm g}}\pd{P_\phi}{T}$\\\vspace{-2.5ex} & \vspace{-2.5ex}\\\hline
       $\displaystyle \ln(E)$ & $\displaystyle\frac{\bar E}{E}\pd{\ln\bar E}{\ln T}+\frac{T}{E}\pd{E_\phi}{\ln T}$\\\vspace{-2.5ex} & \vspace{-2.5ex}\\\hline
       $\displaystyle \ln(s)$ & $\displaystyle \frac{\bar s}{s}\pd{\ln\bar s}{\ln T}+\frac{T}{s}\pd{s_\phi}{\ln T}$\\\vspace{-2.5ex} & \vspace{-2.5ex}\\\hline
       $\displaystyle c_V$ & $\displaystyle\pd{\bar{c}_v}{\ln T}+T\pdd{E_\phi}{T}$\\ \vspace{-2.5ex} & \vspace{-2.5ex}\\\hline
       \multirow{2}{*}{\centering$\displaystyle\chi_\rho$} & $\displaystyle-\chi_T\chi_\rho + \frac{\rho T}{P}\pdt{\bar P}{T}{\rho}$\\ & $\displaystyle\pdt{\bar P}{T}{\rho}=\frac{\bar P}{\rho T}\left(\pd{\bar\chi_T}{\ln\rho}+\bar\chi_T \bar\chi_\rho\right)$\\ \vspace{-2.5ex} & \vspace{-2.5ex}\\\hline
       \multirow{2}{*}{$\displaystyle\chi_T$} & $\displaystyle \chi_T-\chi_T^2+\frac{T^2}{P}\left(\pdd{\bar P}{T}+\pdd{P_\phi}{T}\right)$ \\ & $\displaystyle\pdd{\bar P}{T}=\frac{\bar P}{T^2}\left(\bar\chi_T^2-\bar\chi_T+\pd{\bar\chi_T}{\ln T}\right)$\\\vspace{-2.5ex} & \vspace{-2.5ex}\\\hline
       \multirow{2}{*}{$\displaystyle\Gamma_3$} & $\displaystyle \frac{X}{Y}\left[-1+\frac{1}{\chi_T}\pd{\chi_T}{\ln T}-\frac{1}{c_v}\pd{c_v}{\ln T}+\frac{1}{P}\pd{P}{\ln T}\right]$\\
       & $\displaystyle \pd{P}{\ln T}=\bar{P}_{\rm gas}\pd{\ln \bar{P}_{\rm gas}}{\ln T}+T\pd{P_\phi}{T}+4 P_{\rm rad}$\\\vspace{-2.5ex} & \vspace{-2.5ex}\\\hline
       $\displaystyle\Gamma_1$ & $\displaystyle\pd{\chi_\rho}{\ln T}+\frac{X}{Y}\pd{\chi_T}{\ln T}+\chi_T\pd{\Gamma_3}{\ln  T}$\\\vspace{-2.5ex} & \vspace{-2.5ex}\\\hline
       $\displaystyle\nabla_{\rm ad}$ & $\displaystyle-\frac{\nabla_{\rm ad}}{\Gamma_1}\pd{\Gamma_1}{\ln T}+\frac{1}{\Gamma_1}\pd{\Gamma_3}{\ln T}$\\\vspace{-2.5ex} & \vspace{-2.5ex}\\\hline
       $\displaystyle c_P$ & $\displaystyle\frac{c_v}{\chi_\rho}\pd{\Gamma_1}{\ln T}+\frac{\Gamma_1}{\chi_\rho}\frac{c_v}{\ln T}-\frac{c_P}{\chi_\rho}\pd{\chi_\rho}{\ln T}$\\\vspace{-2.5ex} & \vspace{-2.5ex}\\\hline
       $\displaystyle\left(\pd{s}{T}\right)_\rho$ & $\displaystyle\pd{}{\ln T}\left(\pd{\bar s}{T}\right)+ T \pdd{s_\phi}{T}$\\\vspace{-2.5ex} & \vspace{-2.5ex}\\\hline
       $\displaystyle\left(\pd{s}{\rho}\right)$ & $\displaystyle\pd{}{\ln T}\left(\pd{\bar s}{\rho}\right)+ T \pdt{s_\phi}{T}{\rho}$\\\vspace{-2.5ex} & \vspace{-2.5ex}\\\hline
       $\displaystyle\left(\pd{E}{\rho}\right)$ & $\displaystyle\pd{}{\ln T}\left(\pd{\bar E}{\rho}\right)+ T \pdt{E_\phi}{T}{\rho}$\\\hline

    \end{tabular}
    \caption{Derivatives of the MESA EOS variables listed in table \ref{tab:mesa_EOS} with respect to $\log T$ at constant density ($\pa/\pa\ln T|_\rho$). Overbars denote quantities returned by the default EOS that do not include DM. We have defined the following quantities: $X=P\chi_T$m $Y=\rho T c_v$, $E_\phi=u_\phi/\rho$, and $P_{\rm rad}=aT^4/3$. Some quantities require a second formula for deriving the derivatives of MESA default quantities. If this is the case, the formula is given on the line below. We have derived analytic fitting functions for the derivatives of the DM variables appearing in the final column which we implement into MESA. Note that MESA uses $\ln(P_{\rm g})$, $\ln(E)$, and $\ln(s)$ as variables, hence the logarithms in the first three rows.  }
    \label{tab:derive_const_T}
\end{table*}

\begin{table*}[h]
    \centering
    \begin{tabular}{|c|c|}\hline
        Quantity & Derivative  \\\hline
        $\displaystyle\ln(P_{\rm g})$ & $\displaystyle\frac{\bar{P}_{\rm gas}}{P_{\rm gas}}\left(\frac{\partial \ln \bar{P}_{\rm gas}}{\partial\ln\rho}\right)_T$\\\vspace{-2.5ex} & \vspace{-2.5ex}\\\hline
        $\displaystyle\ln(E)$ & $\displaystyle\frac{\bar E}{E}\frac{\partial \ln \bar{E}}{\partial\ln\rho}+\frac{\rho}{E}\frac{\partial E_\phi}{\partial \rho}$\\\vspace{-2.5ex} & \vspace{-2.5ex}\\\hline
        $\displaystyle\ln (s)$ & $\displaystyle\frac{\bar s}{s}\frac{\partial \ln \bar{s}}{\partial\ln\rho}+\frac{\rho}{s}\frac{\partial s_\phi}{\partial \rho}$\\\vspace{-2.5ex} & \vspace{-2.5ex}\\\hline
        $\displaystyle c_V$ & $\displaystyle \left(\frac{\partial \bar{c}_v}{\partial\ln\rho}\right)_T +\rho\frac{\partial^2 E_\phi}{\partial T \partial\rho}$\\\vspace{-2.5ex} & \vspace{-2.5ex}\\\hline
        \multirow{2}{*}{$\displaystyle\chi_\rho$} & $\displaystyle\chi_\rho-\chi_\rho^2+\frac{\rho^2}{P}\pdd{\bar{P}}{\rho}$
        \\ & $\displaystyle\pdd{\bar P}{\bar\rho}=\frac{\bar P}{\rho^2}\left(\bar\chi_\rho^2-\bar\chi_\rho+\pd{\bar\chi_\rho}{\ln\rho}\right)$\\\vspace{-2.5ex} & \vspace{-2.5ex}\\\hline
        \multirow{2}{*}{$\displaystyle \chi_T$} & $\displaystyle -\chi_T\chi_\rho+\frac{\rho T}{P}\pdt{\bar P}{T}{\rho}$\\ & $\displaystyle\pdt{\bar P}{T}{\rho}=\frac{\bar P}{\rho T}\left(\pd{\bar\chi_T}{\ln\rho}+\bar\chi_T \bar\chi_\rho\right)$\\\vspace{-2.5ex} & \vspace{-2.5ex}\\\hline
        $\displaystyle\Gamma_3$ & $\displaystyle \frac{X}{Y}\left[-1+\frac{1}{\chi_T}\pd{\chi_T}{\ln\rho}-\frac{1}{c_v}\pd{c_v}{\ln\rho}+\frac{\bar{P}_{\rm gas}}{P}\pd{\ln \bar{P}_{\rm gas}}{\ln\rho}\right]$\\\vspace{-2.5ex} & \vspace{-2.5ex}\\\hline
        $\displaystyle\Gamma_1$ & $\displaystyle \pd{\chi_\rho}{\ln\rho}+\frac{X}{Y}\pd{\chi_T}{\ln\rho}+\chi_T\pd{\Gamma_3}{\ln\rho}$\\\vspace{-2.5ex} & \vspace{-2.5ex}\\\hline
        $\displaystyle\nabla_{\rm ad}$ & $\displaystyle-\frac{\nabla_{\rm ad}}{\Gamma_1}\pd{\Gamma_1}{\ln\rho}+\frac{1}{\Gamma_1}\pd{\Gamma_3}{\ln\rho}$ \\\vspace{-2.5ex} & \vspace{-2.5ex}\\\hline
        $\displaystyle c_P$ & $\displaystyle\frac{c_v}{\chi_\rho}\pd{\Gamma_1}{\ln\rho}+\frac{\Gamma_1}{\chi_\rho}\frac{c_v}{\ln\rho}-\frac{c_P}{\chi_\rho}\pd{\chi_\rho}{\ln\rho}$\\\vspace{-2.5ex} & \vspace{-2.5ex}\\\hline
        $\displaystyle \left.\frac{\pa E}{\pa\rho}\right\vert_T$ & $\displaystyle \rho\pdd{\bar E}{\rho}+\rho\pdd{ E_\phi}{\rho}$\\\vspace{-2.5ex} & \vspace{-2.5ex}\\\hline
         $\displaystyle \left.\frac{\pa s}{\pa\rho}\right\vert_T$ & $\displaystyle\rho\pdd{\bar s}{\rho}+\rho\pdd{ s_\phi}{\rho}$\\\vspace{-2.5ex} & \vspace{-2.5ex}\\\hline
          $\displaystyle \left.\frac{\pa s}{\pa T}\right\vert_\rho$ & $\displaystyle \rho\pdt{\bar s}{T}{\rho}+\rho\pdt{ s_\phi}{T}{\rho}$\\\hline
        
    \end{tabular}
    \caption{Derivatives of the MESA EOS variables listed in table \ref{tab:mesa_EOS} with respect to $\log \rho$ at constant temperature ($\pa/\pa\ln \rho|_T$). Overbars denote quantities returned by the default EOS that do not include DM. We have defined the following quantities: $X=P\chi_T$m $Y=\rho T c_v$, $E_\phi=u_\phi/\rho$, and $P_{\rm rad}=aT^4/3$.. Some quantities require a second formula for deriving the derivatives of MESA default quantities. If this is the case, the formula is given on the line below. We have derived analytic fitting functions for the derivatives of the DM variables appearing in the final column which we implement into MESA. Note that MESA uses $\ln(P_{\rm g})$, $\ln(E)$, and $\ln(s)$ as variables, hence the logarithms in the first three rows.  }
    \label{tab:derive_const_d}
\end{table*}

\bibliography{refs}

\end{document}